\documentclass[conference]{IEEEtran}
\IEEEoverridecommandlockouts
\usepackage{cite}
\usepackage{amsmath,amssymb,amsfonts}
\usepackage{algorithm}
\usepackage{algorithmic}
\usepackage{graphicx}
\usepackage{textcomp}
\usepackage{xcolor}
\usepackage{xspace}
\usepackage{soul}
\usepackage{kotex} 

\def\BibTeX{{\rm B\kern-.05em{\sc i\kern-.025em b}\kern-.08em
    T\kern-.1667em\lower.7ex\hbox{E}\kern-.125emX}}
\begin{document}

\title{Trends in Blockchain and Federated Learning for Data Sharing in Distributed Platforms}

\author{\IEEEauthorblockN{Haemin Lee and Joongheon Kim}
\IEEEauthorblockA{Department of Electrical and Computer Engineering, Korea University, Seoul, Republic of Korea 
\\
E-mails: 
\texttt{haemin2@korea.ac.kr},
\texttt{joongheon@korea.ac.kr}
}
}
\maketitle

\begin{abstract}
With the development of communication technologies in 5G networks and the Internet of things (IoT), a massive amount of generated data can improve machine learning (ML) inference through data sharing. However, security and privacy concerns are major obstacles in distributed and wireless networks. In addition, IoT has a limitation on system resources depending on the purpose of services. In addition, a blockchain technology enables secure transactions among participants through consensus algorithms and encryption without a centralized coordinator. In this paper, we first review the federated leaning (FL) and blockchain mechanisms, and then, present a survey on the integration of blockchain and FL for data sharing in industrial, vehicle, and healthcare applications.
\end{abstract}

\section{Introduction}\label{sec:1}
Internet of things (IoT) technologies have been applied in various areas, and thus they generate a massive volume of data. Rapidly increasing IoT network size and accumulated data volume broaden the performance of artificial intelligence (AI) and create new values in various industry applications. 
The collected data should be processed with low latency while assuring security and reliability. Distributed multiple devices can work collaboratively to show better performance~\cite{pieee202105park}. Therefore, effective deployment and intelligent utilization of IoT are needed. However, classical centralized learning algorithms in the IoT are vulnerable to security~\cite{isj2021dao}. 
Centralized machine learning methods which involve data sharing from end devices to the centralized third-party server have privacy and data leakage issues. 
In addition, when data size is large, centralized machine learning algorithms have difficulties in terms of accuracy and efficiency.
with these challenges, federated learning (FL) can be one of alternative solutions that enables on-device machine learning without directly transfer the private end-user data to a central server~\cite{pieee202105park}.

Despite the benefits, there are still remaining issues in FL computation. Since blockchain has been widely used to address security issues in distributed scenarios, recently, there are many approaches to enhance the privacy and security problems by replacing the federated learning server with blockchain~\cite{isj202003saad,isj2021boo}.
In this paper, we first present the existing blockchain and FL frameworks. Then, we present the privacy-preserving data sharing for a distributed industrial, vehicle, and healthcare IoT applications. Finally, we present several open research challenges with their possible solutions.
The rest of the paper is organized as follows. Section~\ref{sec:2} discusses the background of blockchain and federated learning technologies and how blockchain helps in existing FL techniques. Section~\ref{sec:3} describes blockchain applications for AI. Open research challenges and issues are discussed in Section~\ref{sec:4}. Section~\ref{sec:5} concludes the paper.

\section{Related work}\label{sec:2} 

\subsection{Federated Learning (FL)} 
FL is a technique in which multiple local clients, i.e., IoT devices, smartphones, and one central server work together to learn a global model in a decentralized context. Each local client computes an update to the current global model maintained by the server, and only this update is communicated instead of the whole dataset.
The FL is definitely useful in twofold: data privacy and communication efficiency. First, FL has less communication cost as it simply transfers its updates while transferring the local data to a central server put more burden on network traffic and storage costs. Second, the dataset of the local client is never uploaded to the server, which reduces the data leakage. The advantage is decoupling of model training from the need for direct access to the raw training data~\cite{federated}. For applications where the training objective can be specified on the basis of data available on each client, FL can significantly reduce privacy and security risks by limiting the attack surface to only the device, rather than the device and the cloud~\cite{federated}. However, there are still other issues associated with FL. Malicious actors may attempt to compromise the global model or access the client’s dataset. In addition, the lack of incentive or motivation for the client to cooperate in the FL system.
To this end, adopting and leveraging blockchain features for FL applications can be a solution. 

\subsection{Blockchain}
Blockchain is a distributed, open-source, immutable, public digital ledger which is distributed among networked peers. Fundamentally, blockchain is a chain of blocks that make up the ledger with consensus algorithms and encryption. This ledger holds a permanent record of transactions and interactions that took place among participants accessing the distributed and decentralized blockchain network~\cite{blockchain}. This network configuration that consists only of agreed blocks between users is robust to malicious attacks. Consensus protocols such as proof-of-work (PoW), proof-of-steak (PoS) is the method to reach a consensus in a distributed environment that can impact the performance. Blockchain can also be applied to financial services, smart contracts, IoT, and security services. Businesses that require high reliability and honesty can use blockchain to attract customers. Besides, it is distributed and can avoid the single point of failure situation~\cite{overview}. 
With these characteristics, the combination of AI and blockchain technologies has paved the way for many stable systems that support the interaction of multiple devices while providing confidentiality, authentication, and integrity~\cite{blockchain}.

\subsection{Framework}
In the blockchain-based FL (BCFL) framework, the decentralized nature of blockchain can replace the central server~\cite{framework}. The functions of the centralized server can be implemented by the smart contract (SC) instead, and be actuated by transactions on the blockchain. That is, the participating nodes perform FL via blockchain, which maintains the global models and local updates.
The mechanism consists of blockchain storage, committee consensus mechanism, and model training part. The storage of BCFL is a blockchain system that only authorized devices can access the FL training contents. In the committee consensus mechanism, a few honest nodes constitute a committee that validates the updates and assigns a score on them. Only the qualified updates will be packed onto the blockchain. And at the beginning of every round, a new committee is formed. For the model training, the nodes other than committees perform local training each round.

\section{Data Sharing in Distributed Platforms}\label{sec:3}
\begin{figure}
    \centering
    \includegraphics[width =1\columnwidth]{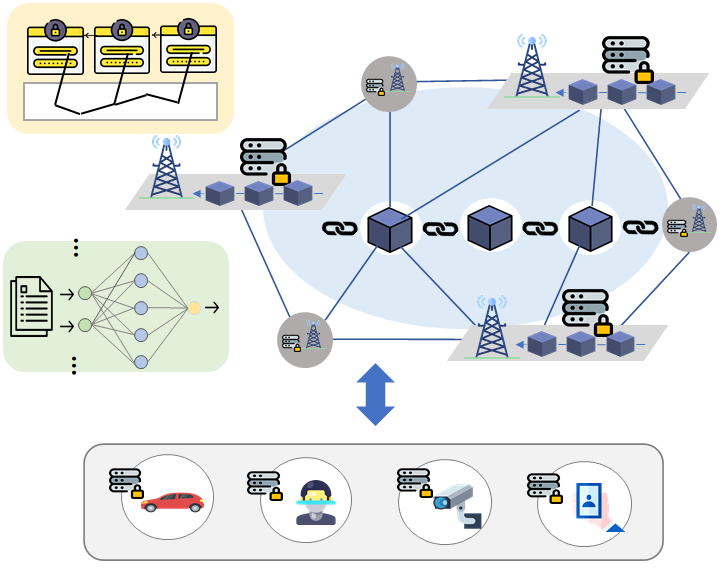}
    \caption{Blockchain Federated learning application}
    \label{fig:1}
\end{figure}

In this section, we describe the integrated blockchain and FL for various applications as Fig.~\ref{fig:1}. The authors in~\cite{IIoT} propose a mechanism for distributed multiple parties in IIoT applications, which incorporates federated learning into permissioned blockchain. In~\cite{vehicle}, a decentralized approach for the vehicle to relieve transmission load and address privacy concerns of providers is proposed. In~\cite{health}, the combination of federated learning and blockchain technology adds progressive value to the healthcare sector.

\subsection{Industrial Internet of Things (IIoT)}
IIoT devices have some different requirements, unlike the consumer IoT that focuses more on user’s comfort and convenience. IIoT is deployed in mission-critical industries and must carry out an advanced task, not simply feeding sensed data to control systems. IIoT-enabled industry requires additional security implementations throughout the operation process, and it must be highly reliable even in harsh conditions. Moreover, effective decision-making is required through sophisticated data analysis. As a sequence, the distributed IIoT can utilize collaborate data sharing to meet the quality of service.
However, in the existing data sharing scheme with the distributed multiple parties has the risk of data leakage. Data leakage may take place during data storage, data transmission, and data sharing, which may lead to severe issues for both owners and providers. 

Authors in~\cite{IIoT}, propose a collaborative architecture that incorporates federated learning into permissioned blockchain to share data to reduce the risk of data leakage. Permissioned blockchain establishes secure connections among all the end IoT devices through its encrypted records, which are maintained by the entities equipped with computing and storage resources.
Different from the existing consensus such as PoW, they applied federated learning empowered consensus, i.e., proof-of-quality (PoQ) which makes better use of the node's computing resources.
%
\subsection{Internet of Vehicles (IoV)} 
With the development of communication technologies in 5G networks and IoT, the automotive industry continues to accelerate towards connected cars~\cite{tvt201905shin,tvt202106jung}. The Internet of Vehicles (IoV) is a distributed network that supports the use of data created by connected cars and vehicular networks. In IoV settings, a massive amount of diverse types of data is constantly generated by the moving vehicles, which includes additional data such as trajectories, traffic information, and multimedia data. 
Through the data sharing, each contributor can make better use of their data by utilizing them together to implement a collaborative task like traffic prediction. However, data sharing in IoV faces two critical challenges~\cite{vehicle}.
First, the vehicles need to share data efficiently despite unreliable inter-vehicle communications. Second, data providers, i.e., IoV are increasingly concerned about data security and privacy issues that can discourage them from providing the data available with them for analysis. 

The presented research results in~\cite{vehicle} integrate blockchain and FL, which can support reliable and secure data sharing among distributed users.
The vehicular networks in~\cite{vehicle} consist of vehicles, road side units (RSUs), macro base stations (MBSs). The MBS first distributes the global model to the blockchain. Then the participating vehicles download the global model from blockchain and train their local models. A certificate authority performs the identification of participants to access the permissioned blockchain. Node selection stage formulates and solves an optimization problem using the deep reinforcement learning (DRL) algorithm to select participating vehicles~\cite{tvt2021jung,twc201912choi}. Then the selected vehicles perform local training on their data by computing the local gradient descent. Vehicles then send the parameters of the trained local model to the nearby RSU and upload it to the blockchain for further verification and aggregation. Then the MBS retrieves the parameters and executes global aggregation.
\subsection{Internet of Health Things (IoHT)} 
With the advent of the Internet of Health Things (IoHT), such as wearable and implantable medical devices, distributedly produced medical data supports real-time analytics and medical tracking, remote monitoring.
However, the Internet of Health Things (IoHT) collects very sensitive health and ambience data that requires privacy. In addition, there are some restrictions on the centralized data collection process.

The presented research results in~\cite{health} present a lightweight hybrid FL framework in blockchain which smart contracts manage the authentication of participating federated nodes, distribution of global or locally trained models. The IoHT interfaces with the edge nodes and the edge nodes have their local, private data for local training and inferencing. The edge nodes are also capable of acting as a local blockchain client or FL worker. 
The communication module of the edge nodes first performs differential privacy and then securely shares the encrypted model and training data to the blockchain App for further processing. 
The blockchain client processes the block creation and shares the smart contract of the hybrid blockchain node for global analysis.

\section{Open Issues}\label{sec:4}
\subsection{FL}
In FL, various communication and networking technologies can be additionally considered in order to improve the performance. First, caching-related technologies can be considered for local training and aggregation~\cite{jsac201806choi,twc201910choi,tmc202106malik,twc202012choi,twc202104choi}. 
In addition, traffic-aware~\cite{isj2021jung} and video/contents-aware~\cite{ton201608kim,tmc201907koo,tmc2021yi} application-specific learning should be additionally considered in order to improve the performance.

\subsection{Blockchain}
In this section, we present the challenges to guarantee data privacy by applying blockchain technique efficiently. First, blockchain itself has security issues such as 51\,\% attack, forking attack and, double spending attack. Moreover, AI-specific consensus protocols can be designed considering proofs based on the quality of learning models or provenance of data~\cite{blockchain}.
Another challenge is the delay and errors of both communications and blockchain on FL. Therefore, some directions need to be taken to minimize the latency.

\section{Conclusions}\label{sec:5}
In this paper, we surveyed and reviewed the privacy-preserving data sharing 
related to the use and applicability of blockchain features for Federated learning. We gave a short overview of blockchain and federated learning and showed how blockchain technology can enhance and solve privacy issues. Moreover, we presented an application of a blockchain federated learning framework in the industrial, vehicle network, and healthcare sectors. Our paper concludes by showing that adopting blockchain for federated learning can definitely leverage the performance of traditional federated learning, but there still exist many research challenges to be addressed.


\section*{Acknowledgment}
This research is supported by the National Research Foundation of Korea (NRF-Korea, 2021R1A4A1030775). J. Kim is a corresponding author of this paper.

\bibliographystyle{IEEEtran}
\bibliography{ref_aimlab,ref_bc}
\end{document}